# A Preliminary Theory for Open Source Ecosystem Micro-economics

**Nicolas Jullien**
**LEGO-M@rsouin**
**IMT Atlantique**
**Nicolas.Jullien@imt-atlantique.fr**

**Klaas-Jan Stol**
**Lero--the Irish Software Research Centre**
**School of Computer Science and Information Technology**
**University College Cork**
**k.stol@cs.ucc.ie**

**James D. Herbsleb**
**Institute for Software Research**
**School of Computer Science**
**Carnegie Mellon University**
**jdh@cs.cmu.edu**

## Introduction

Markets play a key organizing role in most economic systems. Understanding how markets work is critical for effective economic policy. It identifies the levers that policymakers can manipulate to achieve desired effects. Microeconomics uses constructs such as supply and demand, allocation of resources, and equilibria to build models that explain and predict key phenomena such as price setting and the flow of resources to various producers. It allows policy makers to identify market failures, identify abuse of monopoly positions, and other undesirable phenomena and provides theory that points to policy decisions that can have a beneficial impact, minimizing harmful side effects that result in suboptimal outcomes.

Open source ecosystems perform functions analogous to those performed by markets, but they do so without price signals, revenue streams, monetary returns, or other key theoretical mechanisms that are the stock in trade of economists modeling markets. While there has been substantial empirical work identifying factors that influence the contribution to, and use of open source software, we have as yet little theory that identifies the key constructs and relationships that would allow us to explain and predict how open source ecosystems function.

The absence of ecosystem theory is particularly alarming as open source software works its way more broadly and deeply into the economy. As pointed out in a recent report by Eghbal (2016), open source ecosystems are becoming critical digital infrastructure underpinning the publicly and privately produced computational resources we rely on. And it is increasingly apparent that this infrastructure is often neglected and under-resourced, with negative consequences ranging from slowed product development to critical security flaws[1] and propagation of defects and version incompatibilities.

The problem facing policymakers is how to provide support and resources when needed, without distorting decision-making, demotivating volunteers, serving special interests at the expense of others, and maintaining the communities that take on and guide the work. Inappropriate application of resources, for example, could extend the life of a project that should be allowed to decline and be replaced. Adding paid developers to a project could demotivate other volunteers, and reduce the intrinsic motivation of those who are compensated, reducing future contributions. If support is provided to move in a particular technical direction, it could give rise to conflict and potential fragmentation of the community. Interventions that do not respect the logic and underlying principles and relationships of open source ecosystems could easily cause more harm than good, and weaken the very ecosystems it is designed to help.

What is needed is a clearly articulated and empirically validated theory of open source ecosystems. Such a theory should:

- Explain why, how, and when key resources---primarily the work of developers---are attracted to or depart from a project or an ecosystem.

- Explain why, how, and when projects and ecosystems move through a life cycle, from initiation, growth, maturity, and decline and death.

- Explain how decisions about use are made, and how the cumulatively influence the socio-technical position of a project within an ecosystem, and the relations of ecosystems to each other.

The remainder of this chapter provides a sketch of such a theory in the form of a set of propositions, which may form the foundation for future empirical work

# The Three Stages of an Open Source Project

The Stanford economist Paul David identified three factors that influence growth and sustainability of FLOSS projects (David 2006), factors which define three phases in an open source project life: firstly, projects will not be able to enter Phase 2 without achieving sufficient community commitment. In Phase 2, the rate of innovation through addition of new features will ensure growth to Phase 3. In this phase, existential threats emerge through the problem of maintainability, which may be exacerbated by contributor fatigue as key maintainers may leave the project, leaving the project's future in jeopardy.

[1] For example: https://en.wikipedia.org/wiki/Heartbleed

This three-phased idea is a familiar concept in software system development, adoption, and also in project staffing. Software development follows an S-curve process in terms of efficiency, or productivity, called the Rayleigh-Norden curve (Norden 1960). In Phase 1, investments have to be made to develop the foundations of the project, while the production of features might be slow, but the number of people may stay law. After a point, the project enters Phase 2, which is the development phase during which many features are added and the total size of the team may increase. During this phase, the level of productivity tends to be high. After some time, when most of the needs have been addressed, the project enters Phase 3, which is characterized by a decrease in the efficiency of the allocation of resources, and a need to decrease the size of the team affected to the project. Koch (2011) showed that this three-phase evolution in terms of software production applies to open source as well.

As open source has traditionally been a voluntary-based movement, this is not a surprise either as it also echoes the analyses on people's engagement into a collective action (Oliver et al. 1985, Marwell and Oliver, 1993), or an "action taken together by a group of people whose goal is to enhance their status and achieve a common objective" (Wikipedia quoting *Encyclopædia Britannica*). As explained by Marwell and Oliver (1993), motivations and engagement of participants vary, and this explains what happens in these phases in terms of involvement. The first phase attracts only those people who have a high interest in the project, and a low cost of involvement. After some point, and this is especially true for software, increasing returns to adoption start to matter, making more and more interesting to adopt the solution, and attracting new and more diverse actors (Dalle and Jullien 2003, Bonaccorsi and Rossi 2003). Finally, when the project matures, development of new functionality often slows downs and it moves into a mode that could be characterized as "maintenance," which we call Phase 3. Evidence suggests that as projects age, they struggle to recruit and retain newcomers (Von Krogh et al. 2003). This decrease in the growth in participants may simply be the result of, or a signal that, the project has entered a mature phase in which it needs fewer additions and thus fewer contributors (Heckathorn, 1996). Their organization is said to become increasingly bureaucratic (Butler et al, 2008). This is not necessarily a bad thing: as Heckathorn explained (ibid), this bureaucracy makes entry more difficult (more expensive) for newcomers, and thus decrease the number of people willing to participate in such a project.

However, this can lead to the death or downfall of the project. If too many people leave too rapidly, if the project's technical or administrative structure make it increasingly harder to integrate new features, or if, on the contrary, too many people stay for too few things to do, there is an increasing risk of conflicts and inefficient allocation of efforts. Actually, a project's decline, or "death" can occur at any phase of the project, if nobody contributes in the first phase, if growth is not properly managed in Phase 2, or, as stated above, if maturity turns into decay too soon or too quickly.

We detail the ecosystemic challenges of each phase in the remaining sections of this chapter.

## Phase 1: The User-Innovator Phase

The first phase is the one which has attracted probably the most and certainly the earliest research. According to Von Hippel, beyond any motivations, the core of the incentive framework for people to get involved in the early stage of an open source project is the "private collective" innovation model or the "user-as-innovator principle" (Lakhani and von Hippel 2003, von Hippel and von Krogh 2003): as users directly benefit from the innovation they produce, they have an incentive to produce it, and as they can expect add-on, feedback, or cumulative innovation on their own proposition, they have an incentive to freely share it. Jullien and Roudaut (2012) described the difficulties to succeed for projects when the producers were not its users.

As a consequence, to evaluate the chance of success of an open source project in Phase 1, we must focus first on incentives of individual users; developers have to get involved, and, second, the technical and organizational structures that lead them to stay. But first of all, the question is why people or organizations would initiate an open source project.

Early stage FLOSS projects can be classified into three categories. The first category represents the "traditional" FLOSS project, started by one or a few individuals to "scratch an itch" (Raymond 1999). The causes for such itches are various: technical issues that "bother" expert programmers who decide to develop a solution, dissatisfaction with existing proposals, or a lack of existing solutions altogether. Other software solutions may be controlled by a company and lead to market inefficiency (e.g., overpriced products, too little innovation, poor user support, or poor product compatibility). A key characteristic of this type of FLOSS project is that they are solutions developed by individuals to solve a personal computing problem. Examples are widespread, with the Linux kernel perhaps the best known and most successful example. A key challenge for many of these projects is to attract a developer community--most projects have only small developer communities (Comino et al. 2007).

The second category of FLOSS projects is formerly proprietary software that has been open-sourced, such as Netscape's web browser (Ågerfalk and Fitzgerald 2006). The reasons for opensourcing may vary; one reason is that a company no longer wants to spend resources on maintaining the software (van der Linden et al. 2009). Another reason is to increase market share, which will also change the business model around the product (e.g. services around the product). Another reason might be that a company seeks collaboration in development of complementary assets. Here again, one of the key challenges for the opensourcing company is to generate enough interest for the project that it attracts contributors.

The third category is that of so-called "planned" FLOSS projects, typically driven by one or a consortium of companies. One well-known and recent example of this is OpenStack, which was planned by a large consortium of companies, and the goal for the companies involved is to create an industrial standard (or an industrial public good, Romer, 1993).[2] The challenges are those of the creation of an open standard, especially in the balance of the participants' incentive to support the creation of such a standard, and their interest in curving it toward their own goal. For this type of projects, the issue of raising initial investment of resources does not loom large; instead, such projects face organizational challenges such as project governance, which also characterize projects in Phase 2.

The first phase of the model presented (see Fig. S-Curve) illustrates how successful FLOSS projects have an initial stage of growth. Whether or not a FLOSS project will attract sufficient momentum in terms of users and developers (i.e., its popularity) depends on many factors. Firstly, projects in what we have described as Phase 1 attract only those developers who have a very strong interest in the project, or in Raymond's terms (Raymond 2001), those developers that share the same "itch to scratch." These developers typically face low "cost" in participating; for example, they have sufficient time to engage in the project, and they have a significant level of expertise that is required to participate in an early stage of the project when the foundations are laid out. This leads us to pose the following proposition.

> ***Proposition 1:*** *Early-stage FLOSS projects attract developers that perceive the project to be of very high personal value (i.e. it solves a personal problem), and who have low entry barriers to participate (i.e. highly skilled, strong motivation, sufficient time to participate).*

As a FLOSS project is maturing and exhibits a basic feature set beyond the foundations of a project, the project increasingly offers value to stakeholders other than the initial developers who were simply scratching an itch. A more diverse group of stakeholders starts to become interested, including companies who may see business opportunities by leveraging the FLOSS asset for product development, or for developing services around the product. For example, Red Hat is a company (founded in 1993, two years after Linux version 0.0.1 was released) that built an extensive set of services around several very successful FLOSS projects, including the Linux operating system and JBoss. This led Red Hat to become the first one-billion dollar Open Source company in 2012, and its growth has sustained quite significantly in the years since. The process of maturation---that is, the recognition that a project has real potential---may lead to the attraction of additional developers and users, who are perhaps less skilled, or have less spare time available, but who are nevertheless highly motivated to contribute to a project that they are excited about.

---

2The literature on standards is very extensive and well beyond the scope of this chapter. We refer interested readers to Swann's literature review (Swann 2000).

> ***Proposition 2:*** *Early-stage FLOSS projects that offer value beyond "personal interest" will attract a more diverse group of stakeholders than the initial developers.*

There is also a set of socio-technical factors that influence the attraction of new developers to a project, which is a measure of a project's popularity. In terms of technical factors, the implementation technologies may attract, but also deter developers. Modern technologies are typically perceived to be more interesting, not only due to developers' desire and interest to learn those new technologies, but also to improve future job opportunities. Projects that are based on old technologies which are no longer in favor (e.g. Fortran, Cobol) are unlikely to attract today's generation of developers who are likely more interested in modern web technologies such as JavaScript (incl. Node.js) and Python. Thus, we offer the following proposition.

> ***Proposition 3:*** *The popularity of an early-stage FLOSS project depends on the popularity of the technology the project is written in.*

There are few analyses of why companies decide to open source one of its software since Ågerfalk and Fitzgerald (2008) defined the term, as "outsourcing a formally internal software to unknown people" Companies have to adjust in order to be able to collaborate efficiently with communities beyond their organizational boundaries (see, for example, Schaarschmidt et al. 2015). Shaikh and Cornford (2010) have since reaffirmed the fact that embracing an open source strategy means embracing an open source organization, and building trust and cooperative mechanisms with the developers. In exchange, the company, in addition to outsourcing the cost of maintaining and expanding the software can recruit competent and committed developers more easily.

However, there is still a need for a better understanding of the link between internal and external organization, and of the consequences of opening up and collaborate with potential competitors, in other words, to create a sustainable open source ecosystem, or what we refer to here as Phase 2.

> ***Proposition 4:*** *Projects that offer considerable potential business value will attract corporate investment if the project's value proposition is compatible with the company's strategy.*

It is worth noting, first, that not all the open-sourced software projects are aimed at creating value for companies. The traditional outsourcing strategy concerns externalizing complementary assets to specialized companies in order to decrease the total cost of ownership. (We refer to Lacity et al. (2009) for a review of the literature on IT outsourcing).

If a component does not have a great potential to evolve further, it is unlikely to attract any new developers, and, if it is key for the business of the company, it is unlikely to be open-sourced (van der Linden et al. 2009). If there is a potential, it is not sure that the company would want to invest its employees' time and money in developing a community. On the other hand, if a component is business-critical, outsourcing may lead to too much leak toward its competitor (ibid). These simple considerations lead to Table 1, which summarizes the opportunity for outsourcing (and opensourcing).

**Table 1. When outsourcing an internally produced component and how**

|  |  | Central for the core business of the company? | |
|---|---|---|---|
|  |  | **Low importance** | **Business-critical** |
| **Potential of further evolution of the software** | **Low** | Orphan software | Key component, software as a product or service, or internal (closed) maintenance |
|  | **High** | Cooperative development software | solution, open-sourced or not depending on the strategic consequences of open-sourcing the component |

*Proposition 5:* *If a software component is not critical for the core business of a company, and has a potential of evolution, the company will favor an open source strategy to share the cost of development*

*Proposition 6:* *If a software component is critical for the core business of a company, and has a high potential of evolution, an open source strategy will be considered if and only if the technical structure of the software allows the company to keep some strategic components closed while open-sourcing the standard part to benefit from the innovative dynamic of the community,*

As appealing this analysis can be, the case of an opensourcing strategy raises many questions. For example, how can we evaluate the minimum population of skilled developers that is required for the opensourcing strategy to be considered, so as to be able to expect a community to emerge? What does a dynamic of evolution means, is being dynamic enough for a company to prevent competitors from forking and "capturing" the developer community, and the clients? How big should a population of potential clients be for this strategy to be economically sustainable? And, of course, coming back to the howto, for both the core and the complementary asset strategy, how to advertise this open-sourcing to the first contributors, in order to jump-start the ecosystem? What guarantees should a company provide, and how should it structure and publish the software source code to facilitate entry of new developers? What is the cost of sustaining and supporting a community, and what should a company "control" themselves, and which aspects can be left to self-organizing communities, so that project management and leadership can emerge among a core of initial community developers?

To conclude this section, we can say that if the open source licences can be seen, in that perspective, as a new element in the companies' strategic portfolio to manage their relations with a software service provider, it would benefit from more research from fields such as strategic management and information systems regarding the parameters to take into account and the measure of these parameters in the business and financial evaluation of an open source strategy.

If the early-stage software reaches some point of minimum viability, technically and in terms of adoption by a sufficiently large user-base, its adoption by a more general audience may start to grow. To put it in a more global framework, technology adoption, and in Roger's (Rogers 1976) perspective, the end of Phase 1 is characterized by going beyond the core developers team, and even to the early adopters, who if not developing directly, can give feedback, express new needs, but ask also for new support services, such as user-support mailing lists (Kogut and Metiu, 2001) to start reaching the early majority of the 'simple' users.

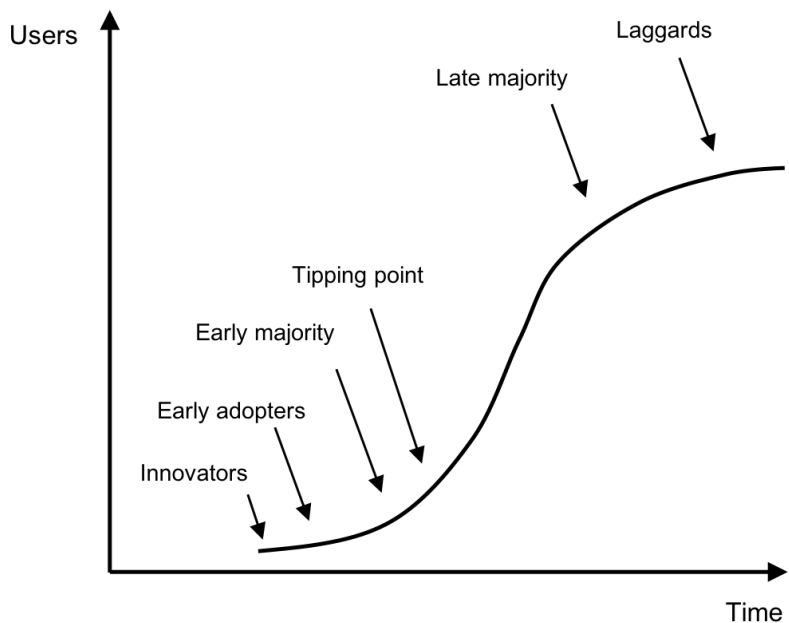

**Figure 4: A Diffusion S-Curve (adapted from Rogers 1976)**

This signals that a project has entered Phase 2, and in this "growth" stage, both the adoption rate and development efforts grow as the utility of the minimally viable product is recognized. We discuss Phase 2 next.

# Phase 2: Blossoming or Fading

As adoption grows, development resources tend to flow into the project, for several distinct reasons. Volunteers are drawn by the increasing visibility and reputation-enhancing potential of contributions to the project. Companies are drawn by the high potential, but not yet fully-realized, of the project for their business--at this relatively early stage, companies may be able to exert some level of control and shape the future of the project. Since the product has demonstrated its utility but is not yet feature-complete, companies invest a portion of their development resources to build the functionality they need.

> ***Proposition 7:*** *Projects that have a sound initial foundation (i.e. the project has commercial potential and represents significant value to users) will attract more developers who may have less time and skills than the original core developers, leading to an increased development velocity of the project.*

As the actors in the community diversify, their goals may diverge, too. As the project grows, the difficulty to maintain a technical coherence may also grow, as the difficulty for newcomers to contribute to the code.

How to deal with the management of stakeholders' different points of view, with the growing complexity of the code and of the organization, how to turn a technical success into a diffusion success, how to make the participation of the companies in the development economically sustainable are some of the main challenges of this phase, and will be discussed this section.

The challenge for projects at that stage is to build the governance structures, the technical infrastructure to allow each to concentrate on their own subject of interest in the project, and everybody to coordinate, so that the diverse range of interests turn into a broader project rather than a battlefield characterized by internal conflicts.

The design of the project into clearly defined modular components is key here, for economic reasons, as it facilitates and decreases the cost of producing new knowledge (Bessen 2005), making entry easier for a new competitor, which, as traditional standard economics pointed out, "needs only to produce a single better component, which can then hook up the market range of complementary components, than if each innovator must develop an entire system" (Farrell 1989). It is also key from a software engineering (Baldwin & Clark 2003) and organizational point of view, as it is very difficult for teams to work efficiently with too many people. In an early study of the Apache web server project, Mockus et al. (2000) hypothesized that open source projects' core teams tend to consist of no more than 15 persons, for accessibility and managerial purpose (Baldwin & Clark 2003).

> ***Proposition 8:*** *Sustainable open source projects are those which succeed in 1) structuring their architecture and their organisation around modules* *managed by small teams; 2) orchestrating the coordination of the differents modules/teams.*

But what exactly characterizes a good open source module team; which are the key qualities of good open source contributors are questions that remain a topic of debate. Team assembly mechanisms can determine team performance (Guimera et al. 2005), especially in creative teams such as those engaged in building knowledge commons (Hess & Ostrom 2006). open source contributor evaluation often relies on the idea of meritocracy, where developers are evaluated based on the quality and quantity of their contributions, which leads to recognition by peers (Jensen & Scacchi 2007). However, meritocratic cultures have been demonstrated to deliver biased observations (Castilla & Bernard 2010), and FLOSS communities have been specifically criticized for this shortcoming (Reagle 2012, Nafus 2012).

In fact, as for other virtual teams (Guimera et al. 2005), social skills in conjunction with leadership behavior affect team motivation and performance, too. Stuart and Gossin (2006) demonstrated how contributors' performance is sensitive to trust and good communication within the team. And, as for any social group, Carillo et al. (2017) insisted on the importance of socialization, i.e. the capacity of the open source organization to teach the rules to newcomers, for them to become good, valuable contributors. Finally, Barcomb et al. (2018), showed that even a limited number of open source project managers may agree on the set of relevant characteristics to identify good open source contributors, they vary in which actual characteristics they use in practice to evaluate different contributors. Even when the different managers use the same attributes, there may be disagreement on the relative importance of these attributes.

*Table 2: relevant characteristics to identify a good open source contributor (from Barcomb et al. 2018)*

|  | **Problematic contributor** | **Good contributor** |
|---|---|---|
| Communication skills (signal over noise ratio) | Too much noise / not enough information | Is good at providing the right level of information |
| Commitment to the project | Unmotivated/passive in seeking answers | Is motivated and does a thorough job |
| Working with others | Tends to find fault with others | Is generally trusting, patient with people |
| Pressure and stress related managing capacity | Gets nervous \ stressed easily (ex.: when things do not go as expected, when there are delays or | Is relaxed, handles stress, technical limitations, setbacks well |
| | due deliverables) | |
| Creativity | Not very creative in terms of solutions | Has an active imagination, proposes creative ideas/solutions |
| Quantity of code contributed | Few lines of code | An impressive quantity of code |
| Quality of code contributed | Tends to provide incomplete or inferior solutions | Produces efficient and well written code, without disturbing other part of the code |
| Global picture: understands the tools / technology / domain, processes behind the project | Low, does not understand beyond the talks/the modules addressed | Understands the technical and non-technical fundamentals of the project |
| Documentation and testing | Does not document/test the code produced, or does so in a way not understandable by others | Documents/test well and clearly the code produced |
| Contribution on other aspects than code (new features, bug description) | Does not contribute beyond code production | Very active in proposing new features, tracking and documenting bugs, etc |

There is a need to develop a better understanding of teaming processes and module-team management as well as ways to articulate project management in such contexts.

> ***Proposition 9:*** *Team composition and skills required may vary according to the technical characteristics and difficulties of the project, but also according to the psychological profile of the team leader.*

> ***Proposition 10:*** *If modularity and delegation of responsibility are key at project level, the organization of this delegation and the level of centralization will vary according to the technical dependencies of the modules, but also according to the psychological and professional profile of the project leader.*

This paradoxical situation in which commercial business relies on the existence and durability of non-market activities questions industrial economics. This is clearly related to "coopetition" questions (Brandenburger and Nalebuff, 1996). As in any cooperative agreement devoted to technology or knowledge development, agents put assets together in a "pre-competitive" phase and share the products of their efforts before coming back to competition (Crémer et al., 1990; Bhattacharya and Guriev, 2006). On the contrary, a FLOSS project is an open game in which the list of players is not bounded ex-ante by a cooperative agreement and whose product is a public good that cannot be privately appropriated by the players. This corresponds closely to the formation of a consortium for the production of a standard.[3]

But there is still a need for a better understanding of the link between open source firms' business models and their investment in the production of open source, when they are at the origin of the project, as said in the previous section, but also when they start contributing to an already existing project. For example, Dahlander and Wallin (2006) showed that firms strategically sponsor individuals who occupy a central position in a community, in order to better access distributed skills and aiming to control the direction of development of the related projects. But not all companies invest so much, and this does not explain why and when companies develop an open source based business model. Based on the concept of "dynamic capabilities" developed by Teece et al. (1997), Jullien and Zimmermann (2011a, 2011b) proposed that when a software project is evolving rapidly in terms of features and development, and when there are sufficiently skilled users to propose contributions, an open source strategy may be valid. The key idea is that a company may be able to propose services based on the *management* of this evolution (support on an official version, ad-hoc developments, and assistance to users, or, a so-called "3A" strategy: Insurance (which spells 'Assurance' in French), Assistance, Adaptation to users' needs). In that case, a company must control the dynamic asset which is the development community--and this requires a deeply involvement in the development of the product as well as in the community. When a product is of less importance to a software company, it may be considered as a complementary asset, and thus, the goal of the company may be to create a consortium to co-develop this component.

> ***Proposition 12:*** *The more central the role of an open source project in a company's business (i.e. a core asset), the more a company will contribute.*
>
> ***Proposition 13:*** *Projects that are "stable" (i.e. little development efforts beyond basic maintenance) tend not to attract corporate investment.*

But how should firms organize themselves to capture the feedback from communities? Ågerfalk and Fitzgerald (2008) observed that to preserve the coexistence and cooperation of two types of organizations that are based on distant albeit not contradictory rationales, firms must, in a nutshell:

- Not seek to dominate and control process
- Provide professional management and business expertise
- Help establish an open and trusted ecosystem.

---

[3] What we mean is that a player offers a standard by developing a software, the other players can adopt and contribute to the development. This "unilateral" adoption is usually called 'bandwagon' in the literature on standards (see, for instance Farrell et Saloner, 1985). See Bessen (2002) and Baldwin and Clark (2003) for a theoretical analysis of the impact of OSS code architecture on the efficiency of libre development. The latter argues that FLOSS may be seen as a new development "institution" (p. 35 and later).

They view such interaction as osmotic rather than parasitic, as the firm's resources reinforce communities' sustainability. But, being able to benefit from the cooperation with an open source project requires internal reorganization, to allow the internal developers to devote a part of their time to these projects, but also to promote cooperative development culture.

As discussed above, companies exert control on open source communities by getting involved in open source communities. Companies do this through sponsorship of selected community members, but they can also do this by having their own developers contribute on open source projects. A key question is how companies can measure the return on investment of such activity, and how can companies manage the involvement of their in-house developers in open source communities? Is such involvement guided by a strategic purpose only (as the employees represent the investment of the firm into project), or are other considerations at play, such as the training of employees, the negotiation of of some compensations (perks) to attract high profile developers? On the other hand, are open source participants using their involvement to signal their high profile to potential employers?

Other questions still are related to legal consequences of 'collective production'. In this context, the rise of open source foundations is a key development. Such foundations are legal entities that represent an open source project. They can also be used as an institutional tool to manage the strategic evolution of a project; one example of this is the OpenStack project.

The projects that succeed in Phase 2 can last for years, and even decades (Linux was first released in 1991 and is still actively developed). From one single project, they expand to other projects and markets, and may even create a whole ecosystem of intertwined projects---the so-called LAMP stack is an example of this (Linux, Apache, MySQL, and Perl/Python/PHP, and today also Ruby). The governance of these projects can become increasingly complex, and some new layers appear to deal with it, and with the multiplicity of projects, such as the foundation system, which can handle the legal representation of the projects, as well as their long term governance.

> ***Proposition 14:*** *projects that become part of a common technology stack will sustain their activity and level of maturity as long as the technology stack as a whole can sustain its activity and level of maturity.*

# Phase 3: Maturity and Beyond

When discussing the maturity phase of open source projects, it is useful to be able to decide whether a project is in fact in its maturity phase. A number of indicators may point to this, for example, a declining or stable number of contributors, contributions, or new features that are added to the project.

> ***Proposition 15:*** *Projects that are stable in terms of number of features added/ removed will lose developers over time as there is a decreasing amount of work left on the project.*

Evidence suggests that as they age, projects find it harder to recruit and retain newcomers (Von Krogh et al. 2003), and their organization is said to become increasingly bureaucratic (Butler et al. 2008). In that respect, these online open projects appear to follow a trend common to traditional organizations, i.e., a natural tendency toward structural inertia when they get bigger, leading to a growing difficulty to adapt (Hannan and Freeman, 1984).

> ***Proposition 16:*** *Mature FLOSS projects tend to become more bureaucratic and rigid in terms of processes and procedures.*

At the same time, as discussed briefly above, the maturity of a project and its ecosystem may suggest that less feature development is needed, which leads to a reduction of the number of involved contributors. While companies may be attracted to new and emerging projects, as they perceive business opportunities the reverse is true as well. Once companies perceive a decline in business value, companies may drop support altogether, for example stopping sponsorship or the support of developers to work on the project.

> ***Proposition 17:*** *Companies that no longer perceive a project to be of business value will stop investing in that project.*

But even among these mature projects, some projects, with the Linux kernel being a prime example (over 25 years old) remain attractive to new developers while others, such as Apache, see decreased participation, but without full demise as some level of maintenance activity is still needed. It remains an open question as to whether this variety is simply due to external dynamics (e.g. technology changes including hardware developments that require projects to constantly adapt itself, as is the case for the Linux kernel)?

> ***Proposition 18:*** *the continuance of external perturbations leads to continued project activity, even when there is no improvement in terms of functionalities.*

Perhaps, are certain governance structures more appropriate or amenable than others? Perhaps certain ecosystem are more resilient; if so, how, and why? Can projects cease due to an increased bureaucracy, and what are some of the consequences for developers and the projects' users? Does formal institutionalization of open source projects (i.e. the creation of foundations) lead to a higher rate of survival?

In other words, how do organizations deal with what Hirschman (1970) called the *exit, voice, and loyalty* phenomenon. When participants in an organization (we consider open source projects as a type of organization) perceive a decrease in quality or benefit to the member, they can either *exit* (withdraw, quit a job, emigrate, stop participating), or they can *voice* (attempt to repair or improve it, express their complaint, or propose changes). The literature stresses the difficulty with the exit strategy in the case of a company, or a country: it is a type of "point of no return" behavior, implying that beyond the fear of losing a job and the salary that comes with it, the fact that employees (or citizens) do not believe in the possible improvement of the situation. Sentimental attachment to the institution may make this belief and the resulting decision to leave even harder. This situation is different for open source projects, because contributors may join and leave the community freely and more easily. Community members could temporarily leave a community during a "cooling down" period. For individual (voluntary) contributors there are no direct consequences, such as the loss of a salary, which means there are lower barriers to the exit strategy, and thus individual contributors may be less willing to negotiate a solution. While contributors' reputation might be at stake (depending on whether they left due to a conflict, for example), for companies coming and going as they please would jeopardize their reputation and credibility significantly; rejoining a community after a company pulled out may be very difficult. When companies that play a key role in an open source community leave, the project's sustainability may be jeopardized.

This analysis could suggest also that:

> ***Proposition 18:*** *A project's core members are the last to abandon a project (they are the most attached to the project), and the peripheral ones the first.*

So, in a nutshell, while in a regular organization (a firm), people may be over loyal (they won't voice when they see a problem, afraid of losing their position), but if they do, they will be very committed to finding a solution, in open source projects, people will probably voice earlier, but also put less effort in finding a solution (and fork or joint a competitive project instead). In the same time, it is not sure that the core members are the best to see the problems and to fix them (to voice). Companies may will voice, but not too much (and possibly not enough), for they may fear to be seen a willing to take the control; they may be also more committed to find a solution, for the project they have invested in to survive

> ***Proposition 18:*** *If a project becomes too bureaucratic while lacking innovation, participants may 'voice,' but those who resent this most are not those who have decision-making power (i.e. core members), or those with business interests (i.e. companies).*

> ***Proposition 19:*** *If a project accepts that it has to reorganize to regain innovativeness, those who have invested the most (core members and companies) will be the most committed to participate in this reorganization.*

However, it is not clear whether this is what happens in reality. Who voice against the slowdown and proposes solution? If the only developers remaining are those hired by companies, will they be sufficiently motivated to sustain a project? Is it wise for companies to stay involved in such projects from a strategic perspective? What might be some indicators that 'predict' such downfall or decline in projects? (Some examples of this could include a decrease in quality or slowdown in bug fixes, etc.) Studies that address contributor behavior, their positions or roles within the project or community, and by drawing careful comparisons with behavior in previous phases may lead to fruitful insights that can help us better understand how to manage these issues.

# Conclusion

Most research on open source software tends to focus on individual software projects, ignoring the complex interactions between the various types of actors listed above, or what is called in this book an open source ecosystem. Open source ecosystems are complex networks of different types of actors at different levels of granularity, including open source projects that rely on other open source projects, companies who either start new, or invest in existing open source projects, open source communities as collections of developers, and of course individual voluntary developers.

Despite two decades of research on open source software, there is very little theory that helps to explain how open source ecosystems "work," evolve, sustain, and decline. There is a considerable body of knowledge on the phenomenon of open source, but much of it is disconnected and has ignored the relationships between different open source projects and between projects and companies. Studies tend to adopt the sample strategy (either developers or projects) or the field study strategy focusing on specific projects, but there is a distinct lack on open source ecosystems that study the *interactions* and *dependencies* between projects. Given the increasing level of interest of companies in open source projects, and also the fact that many companies are built and, indeed, enabled by open source projects, we believe this is a very significant gap in our knowledge base that urgently requires further research, because this will help to better understand the sustainability of open source projects and their entire ecosystems.

In this chapter, we have made an initial attempt to develop such a theory of open source ecosystem "micro-economics," which aims to explain the various forces and behaviors that actors exhibit in open source ecosystems. This initial theory is by no means complete, nor do we have evidence to support our propositions. However, it does help to structure the phenomenon of open source ecosystems, drawing on a three-phased model from the so-called S-curve model, and to formulate propositions regarding where and what is to be studied. This three-phased structure to explain the life cycle of open source projects helps to better understand the chronology of the various challenges that projects face. It also helps to explore the role that companies play in each phase. Furthermore, the structure helps to identify open questions for future research (see Table 3 below).

Finally, the death of a project, and even of an ecosystem, may not be the end of the story (Khondhu et al. 2013). Its technology may survive very long, but it can also generate new ideas, and a part of the developers involved in this former project may use the knowledge they acquired to start something new. For example, the decline of the Geronimo project (a Java/OSGi server runtime environment) seems to have seeded the development of the TomEE project by former Geronimo developers, still within the Apache Foundation projects (Zhou et al. 2016, p. 24).

> ***Proposition 20:*** *Aging projects that suffer from technical and organizational legacy, may be better of being "reinvented" through a new project started from scratch than trying to reorganize the old project.*

**Table 3. Summary of research questions and propositions for future research on open source ecosystems**

*Phase 1: Early Stage*

| Research Questions | Our Propositions |
|---|---|
| How to recruit sufficient and highly skilled developers to ensure successful progress to Stage 2? | Early-stage FLOSS projects attract developers that perceive the project to be of very high personal value (i.e. it solves a personal problem), and who have low entry barriers to participate (i.e. highly skilled, strong motivation, sufficient time to participate). Early-stage FLOSS projects that offer value beyond "personal interest" will attract a more diverse group of stakeholders than the initial developers. The popularity of an early-stage FLOSS project depends on the popularity of the technology the project is written in. Projects that offer considerable potential business value will attract corporate investment if the project's value proposition is compatible with the company's strategy. |
| When and how open-sourcing an | If a software component is not critical for the core business of a company, and has a potential of evolution, the company will |

| in-house software component | favor an open source strategy to share the cost of development

If a software component is critical for the core business of a company, and has a high potential of evolution, an open source strategy will be considered if and only if the technical structure of the software allows the company to keep some strategic components closed while open-sourcing the standard part to benefit from the innovative dynamic of the community, |
|---|---|

*Phase 2: Growth*

| Research Questions | Our Propositions |
|---|---|
| How to design a sustainable project | Sustainable open source projects are those which succeed in<br><br>1. Structuring their architecture and their organisation around modules managed by small teams;<br>2. Orchestrating the coordination of the differents modules/teams. |
| What is an efficient teaming teaming and an efficient management at module level as well as at project level | Team composition and skills required may vary according to<br><br>1. The technical characteristics and difficulties of the project,<br>2. The psychological profile of the team leader<br><br>If modularity and delegation of responsibility is key at project level, the organization of this delegation and the level of centralization will vary according to<br><br>1. The technical dependencies of the modules,<br>2. The psychological and professional profile of the project leader. |
| Corporate investment in open source production | The more central the role of an open source project in a company's business (i.e. a core asset), the more a company will contribute.<br><br>Projects that are "stable" (i.e. little development efforts beyond basic maintenance) tend not to attract corporate investment. |

*Phase 3: Maturity and beyond*

| Research Questions | Our Propositions |
|---|---|
| When does a project enter in a mature phase? | Projects that are stable in terms of number of features added/removed will lose developers over time as there is a decreasing amount of work left on the project.<br><br>Mature FLOSS projects tend to become more bureaucratic and rigid in terms of processes and procedures, making harder for newcomers to get involved.<br><br>The continuance of external perturbations leads to continued project activity, even when there is no improvement in terms of functionalities. |
| Evolution of participation | Companies that no longer perceive a project to be of business value will stop investing in that project.<br><br>A project's core members are the last to leave (they are the most attached to the project), and the peripheral ones the first. |
| Decline or death of a project | If a project becomes too bureaucratic while lacking innovation, participants may 'voice,' but those who resent this most are not those who have decision-making power (i.e. core members), or those with business interests (i.e. companies)<br><br>If a project accepts that it has to reorganize to regain innovativeness, those who have invested the most (core members and companies) will be the most committed to participate in this reorganization.<br><br>Aging projects that suffer from technical and organizational legacy, may be better of being "reinvented" through a new project started from scratch than trying to reorganize the old project. |

**Acknowledgments.** This work was supported, in part, by Science Foundation Ireland grant 15/SIRG/3293 and 13/RC/2094 and co-funded under the European Regional Development Fund through the Southern & Eastern Regional Operational Programme to Lero—the Irish Software Research Centre (www.lero.ie).